\documentclass[runningheads]{llncs}

\usepackage[T1]{fontenc}
\usepackage{graphicx}

\usepackage{booktabs}

\begin{document}

%%
%% The "title" command has an optional parameter,
%% allowing the author to define a "short title" to be used in page headers.
\title{In Defense of Cross-Encoders for Zero-Shot Retrieval}

%%
%% The "author" command and its associated commands are used to define
%% the authors and their affiliations.
%% Of note is the shared affiliation of the first two authors, and the
%% "authornote" and "authornotemark" commands
%% used to denote shared contribution to the research.

\author{Guilherme Rosa\inst{1,2,3} \and Luiz Bonifacio\inst{1,2} \and Vitor Jeronymo\inst{1,2} \and Hugo Abonizio\inst{1,2} \and Marzieh Fadaee\inst{3} \and Roberto Lotufo\inst{1,2} \and Rodrigo Nogueira\inst{1,2,3}}

\institute{NeuralMind, Brazil \and UNICAMP, Brazil \and Zeta Alpha, Netherlands}

\authorrunning{Rosa et al.}
\maketitle              % typeset the header of the contribution

\begin{abstract}
  Bi-encoders and cross-encoders are widely used in many state-of-the-art retrieval pipelines. In this work we study the generalization ability of these two types of architectures on a wide range of parameter count on both in-domain and out-of-domain scenarios.
  We find that the number of parameters and early query-document interactions of cross-encoders play a significant role in the generalization ability of retrieval models. Our experiments show that increasing model size results in marginal gains on in-domain test sets, but much larger gains in new domains never seen during fine-tuning. Furthermore, we show that cross-encoders largely outperform bi-encoders of similar size in several tasks. In the BEIR benchmark, our largest cross-encoder surpasses a state-of-the-art bi-encoder by more than 4 average points. Finally, we show that using bi-encoders as first-stage retrievers provides no gains in comparison to a simpler retriever such as BM25 on out-of-domain tasks. %The code is available at \url{https://github.com/guilhermemr04/scaling-zero-shot-retrieval.git}
  %Recent work has shown that small distilled language models are strong competitors to models that are orders of magnitude larger and slower in a wide range of information retrieval tasks. This has made distilled and dense models, due to latency constraints, the go-to choice for deployment in real-world retrieval applications. In this work, we question this practice by showing that the number of parameters and early query-document interaction play a significant role in the generalization ability of retrieval models. Our experiments show that increasing model size results in marginal gains on in-domain test sets, but much larger gains in new domains never seen during fine-tuning. Furthermore, we show that rerankers largely outperform dense ones of similar size in several tasks. Our largest reranker reaches the state of the art in 12 of the 18 datasets of the Benchmark-IR (BEIR) and surpasses the previous state of the art by 3 average points. Finally, we confirm that in-domain effectiveness is not a good indicator of zero-shot effectiveness. Code is available at \url{https://github.com/guilhermemr04/scaling-zero-shot-retrieval.git}
  \keywords{Cross-encoder, Reranker, Bi-encoder, Dense retrieval, Dual Encoder, Information Retrieval, Zero-shot}
\end{abstract}

\begin{figure}%[ht]
  \centering
  \includegraphics[width=12cm]{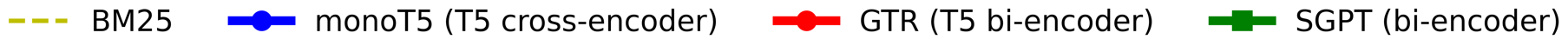}\\
  \includegraphics[width=6cm]{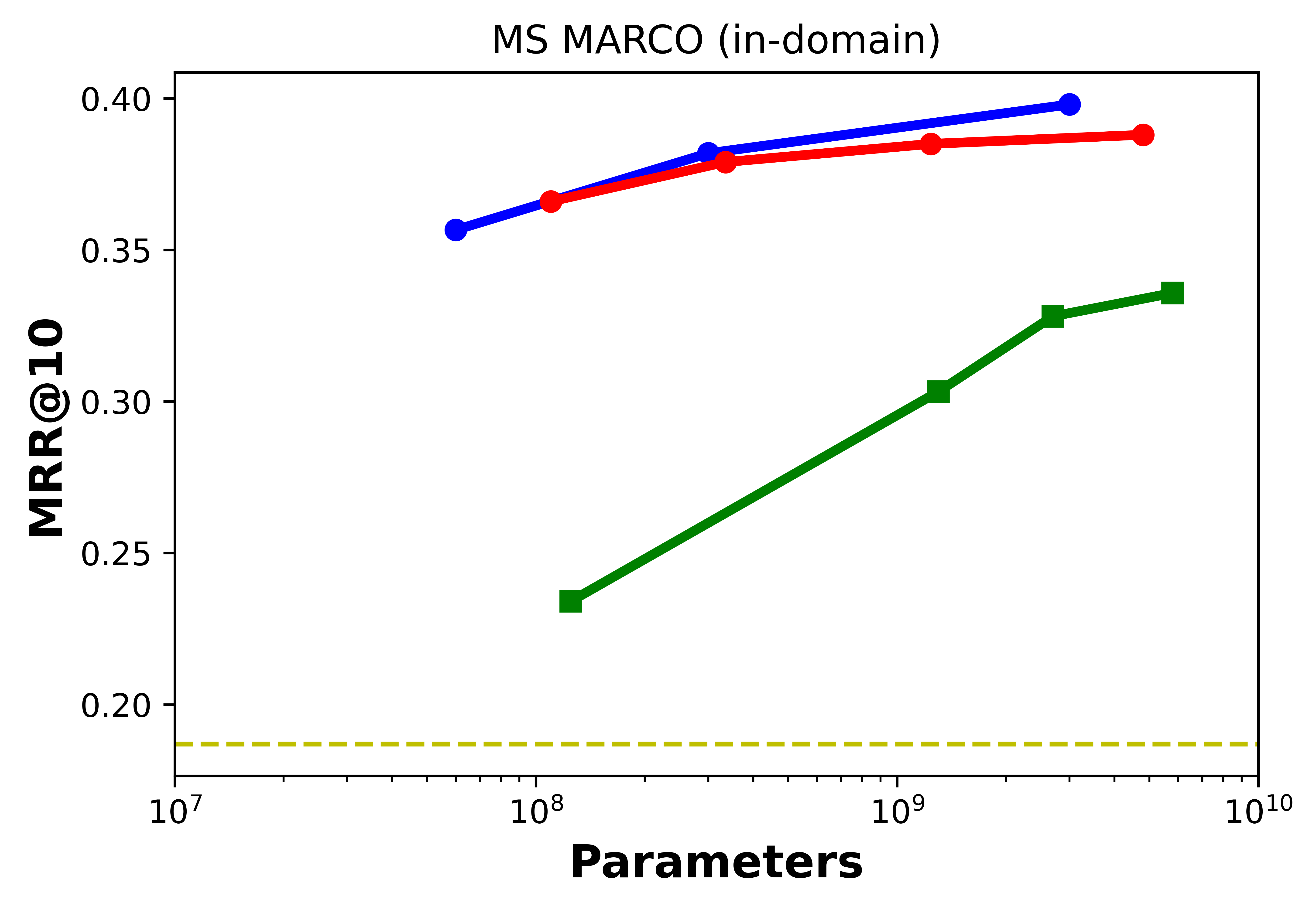}
  \includegraphics[width=6cm]{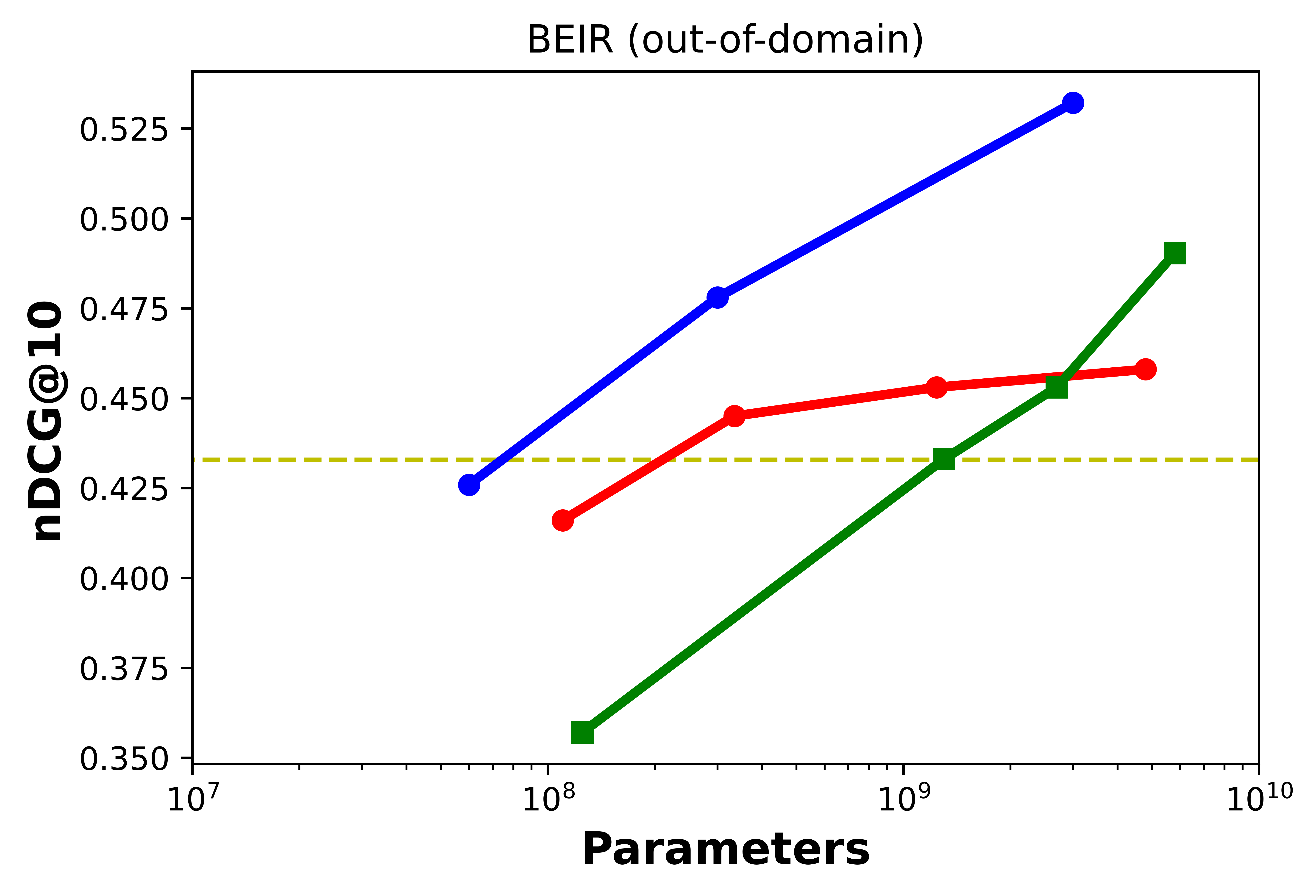}
  \caption{When training and test data are similar, cross-encoders and bi-encoders have similar effectiveness (left). However, cross-encoders outperform bi-encoders of equivalent size on zero-shot tasks (right).}
  \label{fig:graphs_avg}
\end{figure}

\section{Introduction}

The recent success of pretrained language models such as BERT has sparkled the interest in bi-encoder models, generally used as dense retrievers, by the information retrieval community, as they promise to bridge the ``vocabulary mismatch'' problem that has plagued sparse retrievers (e.g., BM25) for decades.
Compared to cross-encoders, which are mostly used as rerankers~\cite{nogueira2019passage,nogueira2020document,macavaney2019cedr,li2020parade,gao2021rethink}, dense retrievers are faster at retrieval time as vector representations of documents can be precomputed prior to retrieval. At retrieval time, only the query vector is computed, and a fast nearest-neighbor search infrastructure is used to retrieve the most similar document vectors to the query vector. Recent work on dense retrieval mostly compares these models with learned sparse representations~\cite{wang2021gpl,xin2021zero,hofstatter2021efficiently,santhanam-etal-2022-colbertv2,lu-etal-2021-less,hofstatter2022introducing,thakur2022domain}, while neural rerankers are often ignored due to their computational costs at retrieval time.

Dense retrievers also show great in-domain effectiveness, i.e., when training and test data are from the same domain. For example, on the MS MARCO leaderboard~\cite{marco} or in the TREC Deep Learning track~\cite{craswell2021trec,craswell2022overview}, many of the top-scoring submissions use dense retrievers as part of their pipeline~\cite{boytsov2020flexible,qiao2020pash,Qiao2022PASHAT,Huang2022YorkUA,gao2022unsupervised,zhang2022hlatr,wu2022contextual}.

However, it is not clear how these retrieval methods compare in zero-shot tasks at different model scales. For instance, recent work shows that the zero-shot effectiveness of dense and sparse retrievers can be improved by distillation or IR-specific pretraining~\cite{santhanam-etal-2022-colbertv2,formal2021spladev2,izacard2021contriever,gao-callan-2022-unsupervised}.
Due to the computational cost, these studies typically use models of the same size as the original BERT, published four years ago. Nevertheless, since hardware and algorithmic improvements reduce the costs of using larger models~\cite{wang2019benchmarking,yoo2022scalable,dettmers2022llm,dettmers2022bit,zeng2022glm}, it is imperative to understand how these different architectures behave at scale. In NLP tasks, for instance, Tay et al.~\cite{tay2022scaling} showed that architectures scale differently and that recently proposed ones do not scale as well as the original Transformer architecture~\cite{vaswani2017attention}.

In this work, we show that dense retrievers and rerankers perform similarly on in-domain tasks, but rerankers largely outperform dense ones across a large range of model sizes.
Our results suggest that current design choices for dense retrieval architectures still have much room for improvement.
Furthermore we show that reranker have similar effectiveness regardless of the initial retrieval algorithm, being it sparse or dense.

\section{Related Work}

Cross-encoders demonstrated state-of-the-art effectiveness in various out-of-domain tasks. For example, Nogueira et al. \cite{nogueira2020document} proposed monoT5, a novel adaptation of a pretrained sequence-to-sequence language model designed for the text ranking task. The model is fine-tuned to generate a score that measures a document's relevance to a given query and has achieved state-of-the-art zero-shot results on TREC 2004 Robust Track \cite{trec2004}. %Furthermore, this approach significantly outperformed fine-tuned models in data scarcity scenarios.
Pradeep et al. \cite{pradeep5h2oloo} used a monoT5 model trained only on a general domain text ranking dataset to reach the best or second best effectiveness on multiple tasks in specific domains, e.g. the medical domain~\cite{Roberts2019OverviewOT}, including documents about COVID-19~\cite{zhang2020rapidly}. Rosa et al. \cite{icail2021}, using the same T5-based model developed for text ranking, showed for the legal domain that a zero-shot model can perform better than models fine-tuned on the task itself. They argued that in a limited annotated data scenario, a zero-shot model fine-tuned only on a large general domain dataset may generalize better on unseen data than models fine-tuned on a small domain-specific dataset.

Although there is extensive study on the effectiveness of IR models on in-domain data, we still know little about their ability to extrapolate to out-of-domain datasets. For instance, Menon et al.~\cite{pmlr-v162-menon22a} argue that with sufficiently large embeddings to represent query and documents, bi-encoders can achieve a similar performance of cross-encoders. However, their analysis is performed only on in-domain tasks.
Zhan et al. \cite{dense2022} found that dense retrieval models can achieve competitive effectiveness compared to reranker models on in-domain tasks, but extrapolate substantially worse to new domains. Their results also indicated that pretraining the model on target domain data improves the extrapolation capability. Ren et al. \cite{drreview} presented an extensive and detailed examination of the zero-shot capability of dense retrieval models to better understand them. The authors analyzed key factors related to zero-shot retrieval effectiveness, such as source dataset, potential bias in target dataset and also reviewed and compared different dense retrieval models.

Ni et al. \cite{dual2021} presented Generalizable T5-based dense Retrievers (GTR), a model designed to address the problem of poor out-of-domain effectiveness of previous bi-encoder models. The authors found that increasing the size of the model brings a significant improvement in its generalization ability. The model outperformed existing sparse and dense models in a variety of retrieval tasks, while being fine-tuned on only 10\% of the MS MARCO dataset. Muennighoff \cite{SGPT} argued that although GPT transformers are the largest language models available, information retrieval tasks are dominated by encoder-only or encoder-decoder transformers. To change this scenario, he presented the SGPT model, a 5.8 billion parameter model that achieved state-of-the-art results on several out-of-domain information retrieval datasets and also outperformed much larger models with up to 175 billion parameters.

\section{Experiments} 

%In our experiments, we evaluate three different models: BM25 \cite{lin2021pyserini}, MiniLM-L6~\cite{minilm} and monoT5~\cite{nogueira2020document} in three different sizes: 60M (small), 220M (base) and 3B parameters.
%%BM25 is a retrieval algorithm that assigns scores to documents based on the terms of a query that appear in it.
%We use BM25 implemented in Pyserini~\cite{lin2021pyserini}, a Python library for information retrieval. MiniLM is a distilled reranker proposed by Wang et al. \cite{minilm} that achieves competitive effectiveness compared to state-of-the-art models in multiple retrieval tasks.
%MonoT5 is an adaptation of the T5 model \cite{raffel2020t5} proposed by Nogueira et al. \cite{nogueira2020document}. The model takes a query-document pair as input and is designed to generate a probability score that quantifies the relevance between them. All three of our transformer models are fine-tuned on MS MARCO \cite{marco} and available at Hugging Face.\footnote{\url{https://huggingface.co/cross-encoder/ms-marco-MiniLM-L-6-v2},\\\url{https://huggingface.co/castorini/monot5-base-msmarco-10k},\\ \url{https://huggingface.co/castorini/monot5-3B-msmarco-10k}} MS MARCO is a large-scale dataset for passage ranking task, consisting of 8.8 million passages taken from the Bing search engine. The fine-tuning procedures we use in this work are described in detail in Thakur et al. \cite{reimers-2019-sentence-bert} and Nogueira et al. \cite{nogueira2020document} for MiniLM and monoT5, respectively. At inference time, rerankers sort a list of 1000 texts retrieved by BM25 according to their relevance to a query.

In our experiments, we compare the performance of three different models: the first is monoT5~\cite{nogueira2020document}, an adaptation of the T5 model \cite{raffel2020t5} proposed by Nogueira et al. \cite{nogueira2020document}. The model takes a query-document pair as input and generates a probability score that quantifies the relevance between them. In our work, the model sorts a list of 1000 texts retrieved by BM25~\cite{lin2021pyserini}. We use monoT5 in three different sizes: 60M (small), 220M (base) and 3B parameters. The second model is GTR, a model available in four different sizes: 110M (base), 335M (large), 1.24B (XL) and 4.8B (XXL) parameters and designed to address the problem of poor out-of-domain effectiveness of previous dense retrieval models.
The third model is the SGPT, the former state-of-the-art model on BEIR, based on the decoder-only architecture of the GPT model. It is also available in four different sizes: 125M, 1.3B, 2.7B and 5.8B parameters.

All three transformer models are fine-tuned on MS MARCO and are available at HuggingFace. MS MARCO is a large-scale dataset for passage ranking consisting of 8.8 million passages taken from the Bing search engine and more than 530k pairs of query and relevant documents available for training. 

We evaluate the zero-shot effectiveness of all models on the BEIR benchmark~\cite{beir}, which consists of 18 publicly available datasets from different domains, such as web, biomedical and financial. With the exception of the MS MARCO dataset, all results are zero-shot, as models are fine-tuned on MS MARCO and directly evaluated on the other datasets.

In our study, we do not evaluate sparse retrievers such as doc2query~\cite{nogueira2019document,nogueira2019doc2query}, DeepCT~\cite{dai2019context}, DeepImpact~\cite{mallia2021learning}, UniCOIL~\cite{lin2021few}, SPARTerm~\cite{bai2020sparterm}, and SPLADE~\cite{formal2021splade} as none of these models is available in larger versions (e.g., with billions of parameters). However, given the competitive results achieved by the SPLADE models~\cite{formal2021spladev2,formal2022spladeplusplus} on the BEIR benchmark, we see scaling up these types of models as a promising direction.
To isolate the effects of cross vs. bi-encoders architectures we also excluded from our analysis models that use IR-specific pretraining (e.g., Contriever~\cite{izacard2021contriever} and CoCondenser~\cite{gao-callan-2022-unsupervised}), distillation (e.g., ColBERTv2~\cite{santhanam-etal-2022-colbertv2} and MiniLM~\cite{minilm}), and dataset augmentation (e.g., InPars~\cite{bonifacio2022inpars} and Promptagator~\cite{dai2022promptagator}). We acknowledge the importance of studying the effect of these methods in zero-shot scenarios but leave it for future work.

\section{Results}

%Results in Table~\ref{tab:main_results} show that while MiniLM achieves excellent results on MS MARCO, including competitive effectiveness to a much larger model such as monoT5-3B, the distilled model underperforms in comparison to monoT5-3B on almost all datasets outside the domain seen during fine-tuning. For example, on the FiQA dataset, which comprises text in the financial domain, MiniLM performs nearly 16 points below our largest reranker. Other significant results include the experiments on the Scifact and Natural Questions datasets, in which the MiniLM is also outperformed by monoT5-3B by almost 10 and 8 points, respectively. Nonetheless, average results show that MiniLM achieves competitive effectiveness compared to monoT5-base, which has 10 times more parameters, suggesting that MiniLM is able to retain some degree of generalization capability.

As shown in Figure~\ref{fig:graphs_avg}, the average results of monoT5-3B and SGPT-5.8B demonstrate that strong zero-shot effectiveness in new text domains can be achieved by increasing the number of model parameters and without fine-tuning on in-domain data. However, our 3 billion parameter reranker model outperformed SGPT, a state-of-the-art dense retrieval model almost twice as large.
Another interesting comparison is that monoT5 and GTR use the same underlying model, T5, but the reranker has better effectiveness than the dense retriever. GTR's effectiveness increases at a slower rate with more parameters in comparison to monoT5's. We suspect that this ``saturation'' is due to the decision to keep the size of document and query vectors fixed to 768 dimensions as the model size increases. We do not observe this behavior for SGPT, as it uses more dimensions as the number of parameters increases. For example, SGPT-125M uses 768 dimensions to represent query and document vector whereas SGPT-5.8B uses 4096 dimensions.
%Results for the MiniLM model reveal a similar behavior: the MiniLM reranker outperforms its dense version on both MS MARCO and BEIR.
These results suggest that early query-document interactions in the reranker architecture and a high number of parameters with equally large number of hidden dimensions contribute positively to zero-shot effectiveness.

\begin{table}
\centering%\centering\resizebox{0.5\textwidth}{!}{
\begin{tabular}{lc|ccc|cccc}
\toprule
 & & \multicolumn{3}{c|}{Rerank Top 1000 BM25} & \multicolumn{4}{c}{Dense Models}\\
 & \textbf{BM25}  & \multicolumn{3}{c|}{\textbf{monoT5}} &  \multicolumn{2}{c}{\textbf{GTR}} & \multicolumn{2}{c}{\textbf{SGPT}}   \\
\textbf{Parameters} & -  & 60M & 220M & 3B & 1.24B & 4.8B & 2.7B & 5.8B \\
\midrule
MS MARCO & 0.187 & 0.356 & 0.381 & \textbf{0.398} & 0.385 & 0.388 & 0.328 & 0.335  \\ 
\midrule
TREC-COVID & 0.594  & 0.692 &  0.777 & 0.794  & 0.584 & 0.501 & 0.807 & \textbf{0.873}  \\
NFCorpus & 0.321  & 0.318 & 0.357 & \textbf{0.383} & 0.343 & 0.342 & 0.339 & 0.363 \\  
BioASQ & 0.522  & 0.488 & 0.524 & \textbf{0.574} & 0.317 & 0.324 & 0.384 & 0.413  \\
%\midrule
Natural Questions & 0.305 & 0.473 & 0.567 & \textbf{0.633} & 0.559 & 0.568 & 0.467 & 0.524 \\
HotpotQA & 0.633 & 0.599 & 0.695 & \textbf{0.759} & 0.591 & 0.599 & 0.528 & 0.593 \\
FEVER & 0.651 & 0.719 & 0.801 & \textbf{0.849} & 0.717 & 0.740 & 0.72 & 0.783 \\
Climate-FEVER & 0.165  & 0.211 & 0.245 & 0.280 & 0.270 & 0.267 & 0.272 & \textbf{0.305} \\
DBPedia & 0.318 & 0.343 & 0.419 & \textbf{0.477} & 0.396 & 0.408 & 0.347 & 0.399  \\
%\midrule
TREC-NEWS & 0.395 & 0.384 & 0.447 & 0.472 & 0.350 & 0.346 & 0.438 & \textbf{0.481} \\
Robust04 & 0.448 & 0.422 & 0.501 & \textbf{0.614} & 0.479 & 0.506 & 0.449 & 0.514 \\
%\midrule
ArguAna & 0.530 & 0.127 & 0.194 & 0.380 & 0.531 & \textbf{0.540} & 0.505 & 0.514 \\
Touché-2020 & \textbf{0.442} & 0.264 & 0.277 & 0.299 & 0.230 & 0.256 & 0.235 & 0.254 \\
%\midrule
CQADupStack & 0.278  & 0.347 & 0.380 & \textbf{0.415} & 0.388 & 0.399 & 0.349 & 0.381 \\
Quora & 0.788 & 0.826 & 0.823 & 0.840 & 0.890 & \textbf{0.892} & 0.856 & 0.846 \\
%\midrule
SciDocs & 0.149 & 0.143 & 0.164 & \textbf{0.197} & 0.159 & 0.161 & 0.165 & \textbf{0.197}  \\
SciFact & 0.678 & 0.696 & 0.735 & \textbf{0.777} & 0.635 & 0.662 & 0.702 & 0.747 \\
%\midrule
FiQA-2018 & 0.236 & 0.337 & 0.413 & \textbf{0.513} & 0.444 & 0.467 & 0.333 & 0.372  \\
%\midrule
Signal-1M (RT) & \textbf{0.330} & 0.271 & 0.277 & 0.314 & 0.268 & 0.273 & 0.249 & 0.267 \\
\midrule
Avg (excl. MARCO) & 0.432 & 0.425 & 0.478 & \textbf{0.532} & 0.453 & 0.458 & 0.453 & 0.490  \\
Improv. over BM25 & - & -0.007 & 0.046 & \textbf{0.100} & 0.021 & 0.026 & 0.021 & 0.058  \\
\bottomrule
\end{tabular}
%}
\vspace{0.1cm}
\caption{Results (nDCG@10) on the BEIR benchmark. We use MRR@10 for MS MARCO. All results except MS MARCO are zero-shot.}
\label{tab:main_results}
\end{table}

\subsection{Combining bi-encoders and cross-encoders}

The results so far showed that cross-encoders have better effectiveness than bi-encoders on out-of-domain tasks. However, the two architectures are not necessarily competitors as they can be combined in a multi-stage pipeline. For example, we can retrieve candidate documents using a bi-encoder and rerank them using a cross-encoder.

In Table~\ref{tab:ablation} we compare the effectiveness (nDCG@10 on BEIR) of using BM25 or GTR-335M to retrieve 1000 documents that are reranked by monoT5-220M. Without a reranker, GTR has a slightly higher nDCG@10 than BM25 (row 1 vs 2). However, there is no advantage in terms of nDCG@10 in using a more expensive dense model over a simple algorithm like the BM25 to provide candidates to the reranker (row 3 vs 4).

%We also experimented with GTR reranking 1000 candidates retrieved by BM25. Since document vectors can be precomputed, this pipeline is much faster than using a cross-encoder reranker. However, in terms of effectiveness, the bi-encoder reranker underperforms in comparison to the cross-encoder (row 4 vs 5).

\begin{table}[]
\centering%\centering\resizebox{0.5\textwidth}{!}{
\begin{tabular}{lll|rrrrrrrr|r}
\toprule
& \textbf{1st stage} & \textbf{Rerank} & \textbf{Covid} & \textbf{NFC} & \textbf{DBP} & \textbf{Touc} & \textbf{Quora} & \textbf{SciD} & \textbf{SciF} & \textbf{FiQA} & \textbf{\ \ Avg.} \\
\midrule
(1) & BM25 & - & 0.594 & 0.321 & 0.318 & 0.442 & 0.788 & 0.149 & 0.678 & 0.236 & 0.441 \\
(2) & GTR & - & 0.557 & 0.329 & 0.391 & 0.219 & 0.890 & 0.158 & 0.639  & 0.424 & 0.451\\
(3) & BM25 & monoT5 & 0.777 & 0.357 & 0.419 & 0.277 & 0.823 & 0.164 & 0.735 & 0.413 & 0.496\\
(4) & GTR & monoT5 & 0.798 & 0.342 & 0.399 & 0.267 & 0.837 & 0.170 & 0.731 & 0.420 & 0.496\\
%(5) & BM25 & GTR & 0.680 & 0.016 & 0.381 & 0.010 &  0.001 & 0.126 & 0.634 & 0.053 & 0.237\\
%BM25+GTR & - & ? & ? & ? & ? & ? & ? & ? & ? & ?\\
%BM25+GTR & monoT5 & ? & ? & ? & ? & ? & ? & ? & ? & ?\\
\bottomrule
\end{tabular}
%}
\vspace{0.1cm}
\caption{nDCG@10 results with different first-stage retrievers (BM25 and GTR-335M). The reranker (monoT5-220M) is fed with 1000 documents retrieved in the first stage.}
\label{tab:ablation}
\end{table}

%\begin{table}[]
%\centering%\centering\resizebox{0.5\textwidth}{!}{
%\begin{tabular}{lcc}
%\toprule
%\textbf{Dataset} & \textbf{BM25      } & \textbf{      GTR-335M} \\
%\midrule
%TREC-COVID & 0.777 & 0.798  \\
%NFCorpus & 0.357 & 0.342  \\
%DBPedia & 0.419 & 0.399  \\
%Touché-2020 & 0.277 & 0.267  \\
%Quora & 0.823 & 0.837  \\
%SCIDOCS & 0.164 & 0.170  \\
%Scifact & 0.735 & 0.731  \\
%FiQA-2018 & 0.413 & 0.420 \\
%\midrule
%Average & 0.496 & 0.496 \\
%\bottomrule
%\end{tabular}
%%}
%\vspace{0.1cm}
%\caption{nDCG@10 results when reranking with MonoT5-220M whose 1000 candidates are retrieved either by BM25 or GTR-335M.}
%\label{tab:ablation}
%\end{table}

\section{Conclusion}

In this work we study how parameter count influences the zero-shot effectiveness of neural retrievers. We begin by showing that in-domain effectiveness, i.e., when retrievers are fine-tuned and evaluated on the same dataset such as MS MARCO, is not a good proxy for zero-shot effectiveness, which corroborates recent claims by Lin et al.~\cite{lin2022fostering}, Gupta et al.~\cite{gupta2022survivorship} and Zhan et al~\cite{dense2022}.

%Furthermore, we show that our monoT5 reranker significantly outperforms the respective similar-size dense retrieval models and its biggest version achieves a new state of the art on most of the datasets used in our zero-shot experiments. This suggests that a large number of parameters may play a significant role in the generalization capability of pretrained language models, since increasing model size improve results, but .

%Furthermore, we show that our largest monoT5 reranker achieves a new state of the art on most of the datasets used in our zero-shot experiments. This suggests that a large number of parameters may play a significant role in the generalization capability of pretrained language models.

Furthermore, our monoT5 rerankers significantly outperform dense retrievers of similar size. This confirms the findings of Zhan et al.~\cite{dense2022}, who show that dense retrievers have poorer generalization ability than rerankers to new domains. Our study, however, shows that this is the case across a wide range of model sizes and text domains, suggesting that much work needs to be done to improve dense retrieval methods.

%Lastly, our results confirm the findings of Zhan et al.~\cite{dense2022}, who show that dense retrievers have poorer generalization than rerankers to new domains. Our study, however, shows that this is the case across a wide range of model sizes and text domains, suggesting that much work needs to be done in dense retrieval methods.

\bibliographystyle{splncs04}
\bibliography{main}

%\pagebreak 

\onecolumn

\section{Appendix}

In this section, we provide graphs that show the zero-shot effectiveness when scaling the parameter count for each dataset on the BEIR benchmark.

\begin{figure*}%[h]
  \centering
  \includegraphics[width=12cm, height=13cm]{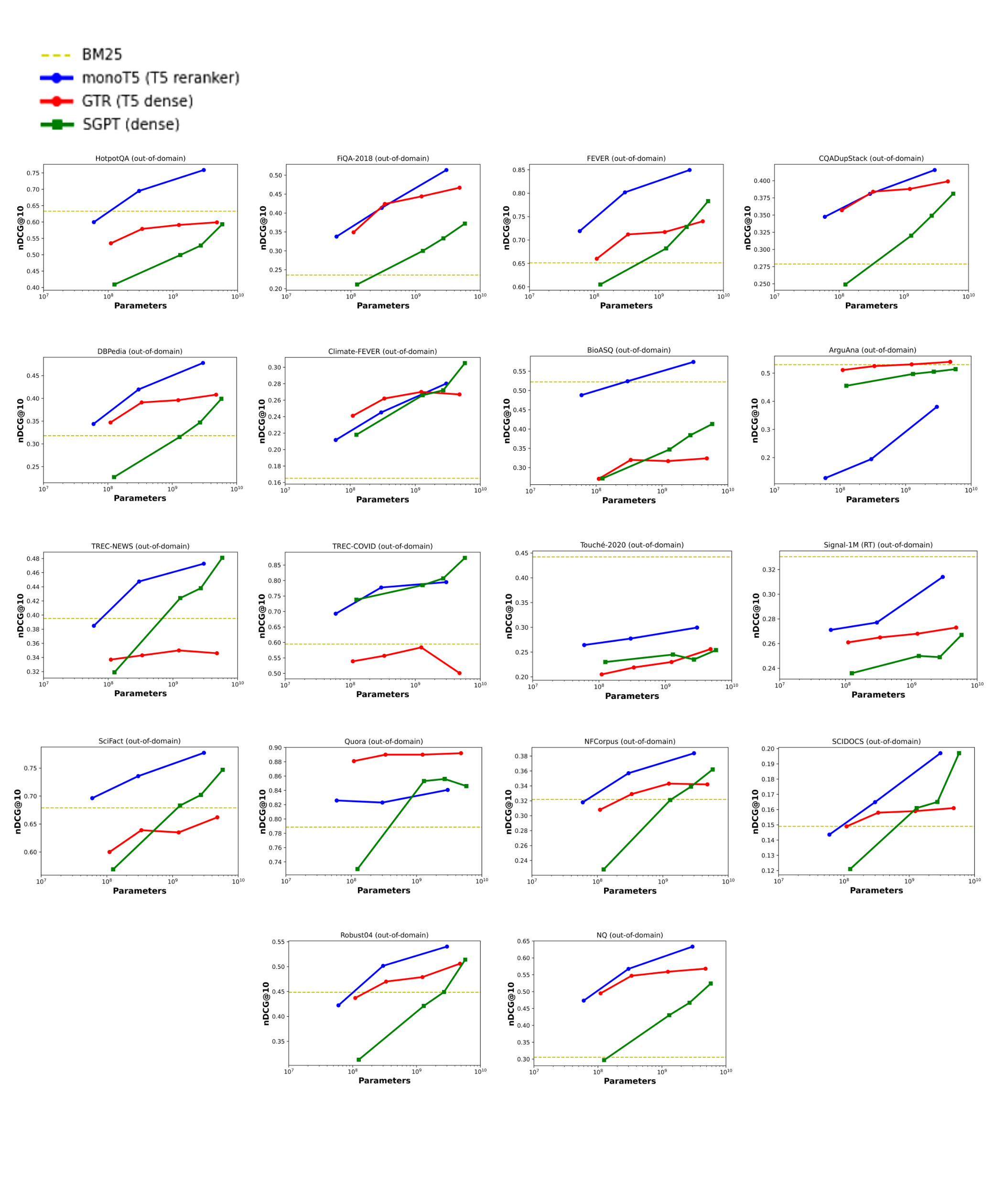}
  \caption{Model size vs effectiveness on out-of-domain data. Effectiveness increase with respect to the number of model parameters in most datasets.}
  \label{fig:graphs}
\end{figure*}

\end{document}